\documentclass[12pt]{article}

\usepackage{graphicx}
\usepackage{a4}

\usepackage{fancybox}
\usepackage{epsfig}
\usepackage{color}

\usepackage{amsfonts}
\usepackage{mathrsfs}
\usepackage{amssymb}
\usepackage{amsmath}

\usepackage{dcolumn}
\usepackage{bm}

\def\pa{\partial}

\def\al{\alpha}
\def\be{\beta}
\def\ga{\gamma}
\def\de{\delta}
\def\ep{\epsilon}

\def\th{\theta}

\def\ka{\kappa}

\def\om{\omega}

\def\Si{\Sigma}

\def\Om{\Omega}

\newcommand{\ben}{\begin{equation}}
\newcommand{\een}{\end{equation}}
\newcommand{\bea}{\begin{eqnarray}}
\newcommand{\eea}{\end{eqnarray}}
\newcommand{\ba}{\begin{array}}
\newcommand{\ea}{\end{array}}
\newcommand{\bit}{\begin{itemize}}
\newcommand{\eit}{\end{itemize}}

\newcommand{\vs}[1]{\vspace{#1 mm}}

\newcommand{\dsl}{\pa \kern-0.5em /}

\begin{document}

\topmargin 0pt \oddsidemargin 0mm

\begin{flushright}

USTC-ICTS/PCFT-22-20 \\

\end{flushright}

\vspace{2mm}

\begin{center}

{\Large \bf Microscopic origin of black hole entropy from tachyon
condensation}

\vs{10}

 {\large Huiquan Li\footnote{E-mail: 32093@qzc.edu.cn}}

\vspace{6mm}

{\em College of Teacher Education, Quzhou University, \\
Quzhou, Zhejiang 324000, China

Yunnan Observatories, Chinese Academy of Sciences, \\
Kunming, Yunnan 650216, China

Peng Huanwu Center for Fundamental Theory, \\
Hefei, Anhui 230026, China}

\end{center}

\vs{9}

\begin{abstract}

We show generically that the dynamics of a probe particle near the
event horizon of a non-extreme black hole is described by the
tachyon effective action. The Hagedorn temperature in the action is
always equal to the Hawking temperature of the background black
hole. The fact suggests that the infalling particle should decay
completely into gravitons or closed strings approaching the event
horizon. The increased area in the black hole due to absorption of a
particle should be interpreted as the entropy of degenerate states
of the closed strings that the particle decays into. With the energy
match condition between the infalling particle and the emitted
closed strings on the event horizon, we examine this variational
area-entropy relation and find that it matches in all cases if the
closed string emission process from an unstable D0-brane obeys the
first law.

\end{abstract}

%\textit{Keywords:} Black hole thermodynamics, entropy, Tachyon condensation

%\textit{PACS:} 11.25.Uv

\section{Introduction}
\label{sec:introduction}
%%%%%%%%%%%%%%%%%%%%%%%%%%%%%%%%%%%%%%%%%%%%%%%%%%%%%%%%%%%%%%%%%%%%%%%%%%%%

The variation of black holes satisfies a set of classical mechanic
laws \cite{Bekenstein:1972tm,Bekenstein:1973ur,Bardeen:1973gs}. The
first law of them tells how the area increases under the change of
black hole quantities, like mass, charge and angular moment:
\begin{equation}\label{e:firstlaw}
 \frac{1}{4}\de A=\frac{2\pi}{\ka}
(\de M-\Phi_H\de Q-\Om_H\de J).
\end{equation}
The close resemblance of the laws to the thermodynamical ones lead
people to realize that thermodynamical properties should be assigned
to black holes, particularly after Hawking found that black holes
could radiate \cite{Hawking:1974sw}. The surface gravity $\ka$ on
the event horizon can be viewed as the Hawking temperature of
radiation: $T_{\textrm{Haw}}=\ka/2\pi$. The area of the black hole
can be viewed as the entropy \cite{Bekenstein:1973ur}:
\begin{equation}\label{e:areentropy}
 S=\frac{1}{4}A.
%\textrm{ }\textrm{ }\textrm{ or }\textrm{ } \textrm{ } \de
%S=\frac{1}{4}\de A.
\end{equation}

A major task since then is to explain this area-entropy relation, as
a means of exploration of quantum gravity. To date, many approaches
have been proposed and adopted to reproduce the relation (e.g., see
reviews \cite{Carlip:2014pma,Perry:2020tts} for a list). But, in
most of the approaches, the microscopic degrees of freedom
responsible for the entropy still remain elusive. This is better
understood for extremal black holes \cite{Strominger:1996sh} whose
near-horizon geometries are AdS space. It is also unknown
how the information of infalling matter that forms the black hole is
recoded in these degrees of freedom.

The unsettled puzzles, together with the information loss paradox,
may imply that some key ingredient is missing in the present black
hole theory. This ingredient is most possibly relevant to the event
horizon, which is usually thought to be smooth for infalling
material. In the fuzzball proposal \cite{Mathur:2005zp}, the
``information-free" horizon is replaced by the boundary of a
fuzzball filled with quantum fuzz. In an alternative proposal
\cite{Almheiri:2012rt}, it is suggested that infalling observer
should encounter firewall at the horizon.

In previous works \cite{Li:2011ypa,Li:2014jfd}, we found that
extraordinary things could really happen naturally near the event
horizon of a non-extreme black hole. In the near-horizon region
whose geometry is generically Rindler space, the action of an
infalling particle is the tachyon field action derived in string
theory. The tachyon effective action for an unstable D0-brane is
\cite{Sen:1999md,Garousi:2000tr,Bergshoeff:2000dq,Kutasov:2003er,
Smedback:2003ur}
\begin{equation}\label{e:tachyonact}
 S_0(T)=-\int d\eta \frac{\tau_0}{\cosh{(\be T)}}
\sqrt{1-(\pa_{\eta}T)^2},
\end{equation}
where the mass of the D0-brane is $\tau_0=2\be/g_s$. The constant
$\be=1/(2l_s)$ for bosonic strings and $=1/(\sqrt{2}l_s)$ for
superstrings, with the string length $l_s=\sqrt{\al'}$. A probe
particle in Rindler space behaves like an unstable D-particle,
described by the action (\ref{e:tachyonact}) at large field $T$.
This means that the particle falling towards a non-extreme black
hole will completely decay into closed strings approaching the
horizon, in terms of the results of tachyon condensation
\cite{Lambert:2003zr}.

If so, the energy of the infalling particle near the event horizon
should be equal to that of the emitted closed strings. The increased
area due to absorption of the particle, which is given by the first
law (\ref{e:firstlaw}), should be accounted for by the entropy of
the closed string states, which is calculated in the tachyon field
theory (\ref{e:tachyonact}): $\de S=\de A/4$. In this work, we
verify this variational area-entropy relation by using the energy
match condition on the event horizon for the three kinds of
black holes in four dimensions: Schwarzschild, RN and Kerr. In contrast
to our previous work
\cite{Li:2011ypa}, we make improvements in two aspects: (i) we show
that the dynamics of the probe particle infalling along more general
geodesics on more general black hole backgrounds can be described by
the tachyon field action; (ii) we use the energy match condition to
estimate the resulting entropy, with no need to consider the
detailed dynamics of each probe particle, which makes a more
universal approach. All through the paper, we adopt the conventions
$G=\hbar=c=k_B=1$.

%%%%%%%%%%%%%%%%%%%%%%%%%%%%%%%%%%%%%%%%%%%%%%%%%%%%%%%%%%%%%%%%%%%%%%%%%%%%
\section{Reissner-Nordstr\"{o}m black hole}
\label{sec:RNBH}
%%%%%%%%%%%%%%%%%%%%%%%%%%%%%%%%%%%%%%%%%%%%%%%%%%%%%%%%%%%%%%%%%%%%%%%%%%%%

The metric of a Reissner-Nordstr\"{o}m (RN) black hole can be
expressed as
\begin{equation}
 ds^2=V(r)(-dt^2+dR^2)+r^2d\Om^2, \textrm{ }\textrm{ }\textrm{ }
V(r)=1-\frac{2M}{r}+\frac{Q^2}{r^2}.
\end{equation}
The constants $M$ and $Q$ are respectively the mass and charge of
the black hole. The condition $V=0$ determines the radii of the
outer and inner horizons: $r_\pm=M\pm\sqrt{M^2-Q^2}$. Here, we
consider the non-extreme black hole with $|Q|<M$. The situation for
Schwarzschild black hole is included as the special case $Q=0$. We
choose the relation between the radial coordinates $R$ and $r$ to
be: $dR=-dr/V(r)$. This gives
\begin{equation}
 R=-(r-r_+-r_-)-\frac{1}{\ep}\left[r_+\ln
\left(\frac{r}{r_+}-1\right)-r_-(1-\ep)\ln
\left(\frac{r}{r_-}-1\right)\right],
\end{equation}
where $\ep=1-r_-/r_+$. Thus, $R\sim-r\rightarrow-\infty$ as
$r\rightarrow\infty$, and $R\rightarrow\infty$ as $r\rightarrow
r_+$. For Schwarzschild black hole, it reduces to be a shift of the
negative tortoise coordinate $r_*$: $R=2M-r_*$. The Hawking
temperature for the black hole is
\begin{equation}\label{e:RNHawtemp}
 T_{\textrm{Haw}}=\frac{\ep}{4\pi r_+}.
\end{equation}

\subsection{Radial infall}

We first consider the case of dropping a neutral particle with mass
$m_0$ and arbitrary radial velocity at infinity. The action of the
particle collapsing towards the black hole along a radial trajectory
is given by
\begin{equation}
 S_0=-m_0\int dt \sqrt{V}\sqrt{1-\dot{R}^2},
\end{equation}
where the dot represents the derivative with respect to the time
coordinate $t$. The energy-momentum tensor from the action is
\begin{equation}\label{e:enmomtensor}
 T_{00}=\frac{m_0\sqrt{V}}{\sqrt{1-\dot{R}^2}}=E,
\textrm{ }\textrm{ }\textrm{ } T_{ij}=-\frac{m_0^2V}{E}\de_{ij}.
\end{equation}
The energy $E$ is constant.

At infinity with $r\rightarrow\infty$, $V\simeq1$ and the action
describes a free particle moving in Minkowski spacetime. The energy
of the particle is $E=\ga m_0$, with $\ga$ being the Lorentz factor.

In terms of the first law (\ref{e:firstlaw}), the surface area of
the black hole will increase when such a particle is eventually
absorbed by the hole. Since the black hole mass increases by an
amount: $\de M=E$, the increased area is
\begin{equation}\label{e:RNincarea}
 \frac{1}{4}\de A=\frac{\de M}{T_{\textrm{Haw}}}
=\frac{4\pi r_+E}{\ep}.
\end{equation}

Let us next focus on the dynamics near the event horizon with
$r\rightarrow r_+$ or $R\rightarrow\infty$. In this region, the
potential $V\simeq\ep e^{-\ep R/r_+}$ and so the action of the
particle is
\begin{equation}\label{e:RNtachact}
 S_0\simeq-m_0\sqrt{\ep}\int dt e^{-\ep R/(2r_+)}\sqrt{1-\dot{R}^2}.
\end{equation}
It is exactly the effective action (\ref{e:tachyonact}) of a rolling
tachyon with $T\rightarrow\infty$ by matching the parameters. This
action arises because the near-horizon geometry is Rindler space
\cite{Li:2011ypa,Li:2014jfd}. It is easy to find that the Hagedorn
temperature in this tachyon action is equal to the Hawking
temperature (\ref{e:RNHawtemp}) of the black hole
\cite{Li:2011ypa,Li:2015qmf}:
\begin{equation}
 T_{\textrm{Hag}}=T_{\textrm{Haw}}.
\end{equation}
From the energy-momentum tensor (\ref{e:enmomtensor}), we know that
the effective theory will evolve into a pressureless ``tachyon
matter" state \cite{Sen:2002in,Sen:2002an} with
$|\dot{R}|\rightarrow1$ and $T_{ij}\rightarrow0$ when approaching
the event horizon.

Hence, the infalling particle near the horizon behaves like an
unstable D0-brane in string theory. The collapsing process towards
the event horizon of a non-extremal black hole should be actually a
tachyon condensation process. The event horizon plays the role of
the closed string vacuum. This implies that the infalling particle
will decay into gravitons or closed strings completely before
reaching the event horizon, via coupling to the gravitational or
closed string modes of the background black hole.

We now estimate the entropy of the closed strings from the decay of
the infalling particle based on the results obtained in
\cite{Lambert:2003zr} (see the Appendix). The results are derived in
the boundary conformal field theory (BCFT), which is an equivalent
description to the tachyon effective theory (\ref{e:tachyonact}).
Therefore, to use the results, we need to first transform the action
(\ref{e:RNtachact}) into the standard ``string-scale" form
(\ref{e:tachyonact}). With the redefinitions $\widetilde{R}=\ep
R/(2\be r_+)$ and $\widetilde{t}=\ep t/(2\be r_+)$, we get
\begin{equation}
 S_0\simeq-\frac{2\be r_+m_0}{\sqrt{\ep}}\int d\widetilde{t}
e^{-\be \widetilde{R}}\sqrt{1-(\pa_{\widetilde{t}}
\widetilde{R})^2}.
\end{equation}
The conserved energy of the infalling particle measured in the new
time coordinate $\widetilde{t}$ is $2\be r_+E/\ep$ (the Hagedorn
temperature measured in this time coordinate is now $\be/2\pi$, the
one in string theory). This energy should be equal to the total
energy $E_\textrm{c}$ of the emitted closed strings. Inserting it in
Eq.\ (\ref{e:energyentropy}) in Appendix, we determine the entropy
of degenerate states of the closed strings emitted from the
infalling particle:
\begin{equation}\label{e:RNradareaentropy}
 \de S=S_\textrm{c}\simeq\frac{4k\pi r_+E}{\ep}=\frac{k}{4}\de A.
\end{equation}
So it is comparable with the increased area (\ref{e:RNincarea}), up
to a numerical constant $k$. When $k=1$, i.e., the closed string
emission process from an unstable D0-brane obeys the first law, the
result is consistent with the expected identification
(\ref{e:areentropy}).

\subsection{Non-radial infall}

For non-radial collapse, the action of the particle is
\begin{equation}\label{e:RNnonradact1}
 S_0=\int dt \mathcal{L}_0=-m_0\int dt \sqrt{V}
\sqrt{1-\dot{R}^2-\frac{r^2\dot{\Om}^2}{V}}.
\end{equation}
The equation of motion for the angular part leads to
\begin{equation}
 -\frac{m_0^2r^2\dot{\Om}}{\mathcal{L}_0}=L,
\end{equation}
where $L$ is a constant. Solving the equation for $\dot{\Om}$ and
inserting the solution into the action (\ref{e:RNnonradact1}), we
get
\begin{equation}\label{e:RNnonradact2}
 \mathcal{L}_0=-\frac{m_0^2r}{\sqrt{L^2+m_0^2r^2}}
\sqrt{V}\sqrt{1-\dot{R}^2}.
\end{equation}

At infinity, the action is also that of a free particle moving in
Minkowski spacetime. We have
\begin{equation}
r\rightarrow\infty: \textrm{ }\textrm{ }\textrm{ }\textrm{ }\textrm{
}\textrm{ } E=\ga m_0, \textrm{ }\textrm{ }\textrm{ } L=\ga
m_0r^2\dot{\Om},
\end{equation}
where $\ga=1/\sqrt{1-\dot{r}^2}$ with $R\sim-r$ towards infinity. So
the constants $E$ and $L$ are respectively the conserved energy and
angular momentum of the particle. Note that the linear velocity
$r\dot{\Om}\rightarrow0$ as $r\rightarrow\infty$ from the latter
equation.

Near the event horizon $R\rightarrow\infty$, the action
(\ref{e:RNnonradact2}) is
\begin{equation}
 S_0\simeq-\frac{\sqrt{\ep}m_0^2r_+}{\sqrt{L^2+m_0^2r_+^2}}
\int dt e^{-\ep R/(2r_+)}\sqrt{1-\dot{R}^2}.
\end{equation}
The action indicates that the particle will decay completely into
closed strings if it can reach the event horizon with appropriate
$L$. With the same procedure, we can find that the relation between
the increased area and entropy is the same as the previous case
(\ref{e:RNradareaentropy}).

\subsection{Charged particle}

The discussion can be extended to the case for a charged particle,
even though we do not know the detailed dynamics. We can do so only
by using the energy match condition on the event horizon.

When the RN black hole absorbs a particle with charge $q$, its
charge increases with $\de Q=q$. Thus, in terms of the first law
(\ref{e:firstlaw}), we only need to replace $E$ in Eq.\
(\ref{e:RNincarea}) by $E-q\Phi_H$, where the surface electric
potential $\Phi_H=Q/r_+$.

In the field theory description, it can be found that the energy of
the charged particle will become $E-q\Phi_H$ when it reaches the
event horizon because work will be done on the charge as it falls in
the electric field of the black hole (the sign of the work depends
on the relative sign between $Q$ and $q$). Approaching the horizon,
the charged particle should also decay completely into closed
strings. In terms of the energy match condition on the event
horizon, the energy of the radiated closed strings is therefore:
$E_\textrm{c}=2\be r_+(E-q\Phi_H)/\ep$. Thus, we get the same
variational area-entropy relation: $\de S\simeq k\de A/4$.

%%%%%%%%%%%%%%%%%%%%%%%%%%%%%%%%%%%%%%%%%%%%%%%%%%%%%%%%%%%%%%%%%%%%%%%%%%%%
\section{Kerr black hole}
\label{sec:KerrBH}
%%%%%%%%%%%%%%%%%%%%%%%%%%%%%%%%%%%%%%%%%%%%%%%%%%%%%%%%%%%%%%%%%%%%%%%%%%%%

The metric for a Kerr black hole with mass $M$ and angular momentum
$J$ is
\begin{equation}
 ds^2=-\frac{\rho^2\triangle}{\Si}dt^2+\frac{\Si\sin^2\th}{\rho^2}
(d\phi-\om dt)^2+\frac{\rho^2}{\triangle}dr^2+\rho^2d\th^2,
\end{equation}
where $\rho^2=r^2+a^2\cos^2\th$, $\triangle=r^2-2Mr+a^2$,
$\Si=(r^2+a^2)^2-a^2\triangle\sin^2\th$ and $\om=2Mar/\Si$ with the
spin parameter $a=J/M$. The radii of the horizons are:
$r_\pm=M\pm\sqrt{M^2-a^2}$. On the event horizon, the angular
velocity of the hole is $\Om_H=\om(r_+)=a/(2Mr_+)$. The Hawking
temperature for the non-extreme Kerr black hole is
\begin{equation}\label{e:Kerrtemp}
 T_{\textrm{Haw}}=\frac{\ep}{8\pi M},
\end{equation}
where $\ep=1-r_-/r_+$.

The Lagrangian of a probe particle moving in the spacetime is also
given by
$\mathcal{L}_0=-m_0\sqrt{g_{\mu\nu}\dot{x}^\mu\dot{x}^\nu}$. The
equation of motion about $\dot{\phi}$ leads to
\begin{equation}\label{e:Kerrangeof}
 -\frac{m_0^2\Si\sin^2\th(\dot{\phi}-\om)}{\rho^2\mathcal{L}_0}=L,
\end{equation}
where $L$ is a constant. The case $L=0$ corresponds to the
trajectory of the zero angular momentum observer (ZAMO). With the
solution to the equation, the Lagrangian becomes
\begin{equation}\label{e:Kerract}
 \mathcal{L}_0=-\frac{m_0^2\rho\sin\th\sqrt{\triangle}}
{\sqrt{\rho^2L^2+m_0^2\Si\sin^2\th}}\sqrt{1-\frac{\Si}
{\triangle^2}\left(\dot{r}^2+\triangle\dot{\th}^2\right)}.
\end{equation}
Towards the event horizon with $\triangle=0$, the angular motion
along the $\th$ direction is suppressed. We only consider the
special case of geodesics with constant angle $\th$ in what follows.
This is allowed at least for geodesics on the equatorial plane in
terms of the Carter constant of motion \cite{Carter:1968rr}.

At infinity, the conserved energy and angular momentum of the
particle are respectively
\begin{equation}\label{e:Kerrinfenmom}
 r\rightarrow\infty: \textrm{ }\textrm{ }\textrm{ }\textrm{ }\textrm{
}\textrm{ } E=\ga m_0, \textrm{ }\textrm{ }\textrm{ } L=\ga
m_0r^2\sin^2\th\dot{\phi}.
\end{equation}
When the particle is absorbed, the mass and angular momentum of the
black hole increase respectively by: $\de M=E$ and $\de J=L$. In
this case, the first law (\ref{e:firstlaw}) reads
\begin{equation}
 \frac{1}{4}\de A=\frac{8\pi M}{\ep}(E-\Om_HL).
\end{equation}

Near the event horizon $r\rightarrow r_+$, the Lagrangian
(\ref{e:Kerract}) with $\dot{\th}=0$ is approximately
\begin{equation}\label{e:Kerrtacact}
 \mathcal{L}_0\simeq-\frac{m_0^2r_+\rho_+\sin\th\sqrt{\ep}}
{\sqrt{\rho_+^2L^2+4m_0^2M^2\sin^2\th}}e^{-\ep
R/4M}\sqrt{1-\dot{R}},
\end{equation}
where $\rho^2_+=1+(a/r_+)^2\cos^2\th$. The $R$ coordinate is defined
by $dR=-\sqrt{\Si}dr/\triangle$. We can also see that the Hagedorn
temperature $T_{\textrm{Hag}}$ in the action is equal to the Hawking
temperature (\ref{e:Kerrtemp}). The tachyon action suggests that the
particle should decay into closed strings completely if it can reach
the event horizon. Its energy should be equal to the total energy of
the emitted closed strings on the event horizon.

Since the black hole spacetime is rotating, we need to choose a
fiducial frame to make the energy match. The fiducial frame can be
chosen as that of a ZAMO \cite{Thorne:1986iy}. At infinity, the
observer is static. The measured conserved energy and angular
momentum are respectively $p_t=-E$ and $p_\phi=L$, as given in Eq.\
(\ref{e:Kerrinfenmom}). On the event horizon, the observer is
co-rotating with the horizon at $\Om_H$, so is the infalling
particle, i.e., $\dot{\phi}\rightarrow\Om_H$ regardless of the value
of $L$ as $r\rightarrow r_+$ in Eq.\ (\ref{e:Kerrangeof}). Thus, it
is natural to think that the emitted closed strings near the horizon
should also be co-rotating with the horizon (in the co-rotating
frame, the particle falls radially, like the previous case).
Generally for an observer rotating with $\Om_H$, the measured
conserved energy becomes $-p\cdot\xi=E-\Om_H L$, where
$\xi^\mu=\pa_t+\Om_H\pa_\phi$ is the co-rotation Killing vector that
is normal to the horizon. This is the energy to match with the total
energy of the emitted closed strings on the horizon.

Similarly, to make the match on the same scale, we need to transform
the action (\ref{e:Kerrtacact}) to the string-scale form
(\ref{e:tachyonact}) with the redefinitions $\widetilde{R}=\ep
R/(4\be M)$ and $\widetilde{t}=\ep t/(4\be M)$. Under the
transformations, the quantity $\Om_HL$ transforms as $E$. In the
redefined coordinates $(\widetilde{t},\widetilde{R})$, the energy
match condition on the event horizon is thus: $E_\textrm{c}=4\be
M(E-\Om_HL)/\ep$. Inserting it into Eq.\ (\ref{e:energyentropy}), we
get the entropy of the closed string states:
\begin{equation}
 \de S\simeq\frac{8k\pi M}{\ep}(E-\Om_HL)=\frac{k}{4}\de A.
\end{equation}

%%%%%%%%%%%%%%%%%%%%%%%%%%%%%%%%%%%%%%%%%%%%%%%%%%%%%%%%%%%%%%%%%%%%%%%%%%%%
\section{Discussion}
\label{sec:discussion}
%%%%%%%%%%%%%%%%%%%%%%%%%%%%%%%%%%%%%%%%%%%%%%%%%%%%%%%%%%%%%%%%%%%%%%%%%%%%

We show that infalling matter trying to enter into a non-extreme
black hole will encounter a tachyon condensation process near the
event horizon, with the Hagedorn temperature equal to the Hawking
temperature. It will decay into closed strings completely before
reaching the event horizon, via coupling to the background
gravitational or closed string modes. So this provides a description
of the infall process beyond the semiclassical approximation and is
obscure in the usual geodesic equation approach.

The revealed fact implies that the black hole area or entropy should
be interpreted as the degeneracy of discrete states of the closed
strings that the infalling matter decays into. All the information
of the matter is recorded in the emitted closed strings. The
information of each infalling particle is mainly recorded at the
closed string state at some cut-off level which is relevant to
energy of the particle on the horizon.

This is a picture for an exterior observer. For an observer
co-moving with the infalling matter, the matter still possibly
enters into the black hole, with nothing special happening near the
horizon, if it encounters no firewall \cite{Almheiri:2012rt}. The
co-moving observer sees Minkowski spacetime there, if the
equivalence principle still holds. The states on the infalling mater
are not excited and the coupling to gravitational modes becomes
trivial.

The field theory description here can not predict the fate and the
final state of the emitted closed strings. They may cross the
horizon and enter into the black hole. In this sense, the black hole
looks like a condensate of gravitons
\cite{Dvali:2011aa,Dvali:2012rt,Alfaro:2016iqs}. But we prefer a
membrane paradigm \cite{Thorne:1986iy}: the information of an
infalling particle will be left on the membrane above the event
horizon once the particle disappears from our world. The surrogates
of the particle on the membrane are the emitted closed strings.
Hawking radiation of a particle can be viewed as the inverse process
of tachyon condensation from a collective of closed strings on the
membrane. Hence, for an exterior observer, the information of the
infalling matter is reformulated in random ways in Hawking
radiation, but is retrievable.

\section*{Acknowledgements\markboth{Acknowledgements}{Acknowledgements}}

We are grateful for helpful discussions with Jianxin Lu. The visit
to PCFT is financially supported by NSFC 12247103.

\appendix

\section*{Appendix}

\setcounter{equation}{0}

\section{Entropy of tachyon condensation}

\renewcommand{\theequation}{A.\arabic{equation}}

The closed string emission from unstable D-branes is studied in the
BCFT in \cite{Okuda:2002yd,Lambert:2003zr}. The tachyon effective
theory (\ref{e:tachyonact}) is an equivalent description to this
conformal theory \cite{Okuyama:2003wm,Lambert:2003zr,Sen:2004nf}.
Thus, the results should be common for the two equivalent theories.
In this Appendix, we summarize the results for the decay from an
unstable D0-brane in bosonic string theory and estimate the entropy
of degenerate states of the emitted closed strings, mainly based on
\cite{Lambert:2003zr,Sen:2004nf}.

The tachyon effective action (\ref{e:tachyonact}) or the BCFT action
coupling to the closed string modes will lead to closed string
emission from the rolling tachyon. The total energy of the emitted
closed strings, up to some numerical factor, is given by:
\begin{equation}\label{e:closeden}
 E_{\textrm{c}}\sim \sum_nd_n\int\frac{d^{25}p_\bot}
{(2\pi)^{23}}e^{-\frac{\pi}{\be} E^{(n)}}.
\end{equation}
The number of states of the emitted closed strings with left-right
symmetry is the same as that of free open strings for large $n$:
\begin{equation}
 d_n\sim n^{-\frac{27}{4}}e^{4\pi\sqrt{n}}.
\end{equation}
The energy $E^{(n)}$ at large level $n$ can be approximated to be
\begin{equation}\label{e:levelen}
 E^{(n)}=\sqrt{M_n^2+p_\bot^2}\simeq M_n+\frac{p_\bot^2}{2M_n}
+\cdots,
\end{equation}
where $M_n=4\be\sqrt{n-1} \simeq 4 \beta n^{1/2}$ for bosonic string
theory and $p_\bot$ is the transverse momentum.  Note that  the most
important contribution  of $p_{\bot}$ to the integral
(\ref{e:closeden}) occurs around $|p_{\bot}/\beta|\sim n^{1/4}\ll
n^{1/2}$ for large $n$, therefore justifying the above approximation
in (\ref{e:levelen}).

Evaluating the integral in (\ref{e:closeden})  gives
\begin{equation}
 E_{\textrm{c}}\simeq\frac{\be}{k}\sum_n\frac{1}{\sqrt{n}},
\end{equation}
where $k$ is an undetermined numerical constant \cite{Sen:2004nf}.
The sum is divergent. Since the energy of D0-brane is finite, this
divergence cannot be true and is due to our lack of consideration of
the backreaction of the closed string emission process during the
rolling of the tachyon.  So a natural cut-off should be chosen and
this should be  on the order of $1/g_{s}$, due to the D0-brane
energy of order $1/g_{s}$.  So this natural cut-off can be large for
a weak string coupling. Denote the cut-off level as $N\sim1/g_s^2$.
For large $n$ (in this case, we can write the sum as an integral
approximately), we have
\begin{equation}
 E_{\textrm{c}}\simeq\frac{2\be}{k}\sqrt{N}
\simeq\frac{1}{2k}M_N.
\end{equation}
Most of the energy $E_{\textrm{c}}$ is stored in closed stings at
level $N\sim 1/g^{2}_{s}$ since any large cut-off $N\ll 1/g^{2}_{s}$
will give an energy $E_{c}\sim \beta \sqrt{N} \ll 1/g_{s}$, much
less than the D0-brane energy. The number of states at this level is
also the largest. Thus, the entropy of the emitted closed strings
can be estimated to be
\begin{equation}\label{e:energyentropy}
 S_\textrm{c}\simeq\ln d_N\simeq\frac{2k\pi}{\be}
E_\textrm{c}.
\end{equation}
Obviously, this tachyon condensation process obeys the first law for
$k=1$. When most energy is converted to the highest level $N$, the
emission process tends to attain the maximum entropy.

For an unstable D0-brane, the energy $E_\textrm{c}$ is the scale of
its mass $\tau_0$. For our case here, it is completely sourced by
the energy of the infalling particle near the event horizon since
the infalling particle decays completely into close strings there,
as indicated by the near-horizon tachyon effective actions.

\newpage
\bibliographystyle{JHEP}
\bibliography{b}

\end{document}